\documentclass[%
 reprint,
 amsmath,amssymb,
 aps,
]{revtex4-1}

\usepackage{graphicx}
\usepackage{dcolumn}
\usepackage{bm}
\usepackage{epstopdf}

\begin{document}

\preprint{APS/123-QED}

\title{Hydrogen atom in a quantum plasma environment under the influence of Aharonov-Bohm flux and electric and magnetic fields}

\author{Babatunde J. Falaye}
\email{fbjames11@physicist.net; babatunde.falaye@fulafia.edu.ng}
\affiliation{Departamento de F\'isica, Escuela Superior de F\'isica y Matem\'aticas, Instituto Polit\'ecnico Nacional, Edificio 9, Unidad Profesional Adolfo L\'opez Mateos, Mexico D.F. 07738, Mexico}
\affiliation{Applied Theoretical Physics Division, Department of Physics, Federal University Lafia,  P. M. B. 146, Lafia, Nigeria}
\author{Guo-Hua Sun}
\email{sunghdb@yahoo.com}
\affiliation{Catedr\'atica CONACyT, CIC, Instituto Polit\'{e}cnico Nacional, Unidad Profesional Adolfo L\'opez Mateos, Mexico D. F. 07700, Mexico}
\author{Ram\'{o}n Silva-Ortigoza}
\email{rsilvao@ipn.mx}
\author{Shi-Hai Dong}
\email{dongsh2@yahoo.com}
\affiliation{CIDETEC, Instituto Polit\'{e}cnico Nacional, Unidad Profesional Adolfo L\'opez Mateos, M\'{e}xico D. F. 07700, M\'{e}xico}%

\author{\bf Phys. Rev. E 93, 053201 – Published 2 May 2016}

\begin{abstract}
This study presents the confinement influences of Aharonov-Bohm-flux (AB-flux), electric and magnetic fields directed along $z$-axis and encircled by quantum plasmas, on the hydrogen atom. The all-inclusive effects result to a strongly attractive system while the localizations of quantum levels change and the eigenvalues decrease. We find that, the combined effect of the fields is stronger than  solitary effect and consequently, there is a substantial shift in the bound state energy of the system. We also find that to perpetuate a low-energy medium for hydrogen atom in quantum plasmas, strong electric field and weak magnetic field are required, where AB-flux field can be used as a regulator. The application of perturbation technique utilized in this paper is not restricted to plasma physics, it can also be applied in molecular physics.

\pacs{34.50.Fa, 52.25.Kn, 52.27.Gr, 52.27.Lw}
\keywords{ Quantum plasmas; Perturbation technique; Hydrogen atom.}
\end{abstract}
\maketitle


\section{Introduction}
\label{sec1}
Plasma is one of the four fundamental states of matter. It is a gas, but not a usual kind of gas. This is because in a normal gas, the atoms or molecules are electrically neutral, but in a plasma at least some of these particles have either lost or gained an electron, so that a plasma consists of free electrons and positively or negatively charged atoms and molecules known as ions. Thus, we can describe a plasma as an ionized gas, a gas into which sufficient energy is provided to free electrons from atoms or molecules and to allow both species, i.e., ions and electrons, to coexist. Moreover, to transform a normal gas into plasma, a very high temperature is required.

The Debye-H\"uckel theory, proposed by authors \cite{BJ1} gave a theoretical explanation for departures from ideality in solutions of electrolytes and plasmas. The Debye-H\"uckel model provides a modern treatment of non-ideality in plasma via the screening effect. This model is use to simulate plasma screening effect of weakly coupled plasmas and it is given by \cite{BJ2}: $V(r)=-(Z e^2/r)\exp(-r/\lambda_D)$,  where $\lambda_D$ represents Debye length or  Debye screening parameter and it determine the interaction between electron-electron in Debye plasma. This model generally accounts for pair correlations. It can be observed from  the model that, the effect of plasmas on a test charge is just a replacement of Coulomb potential by an effective screened potential.  It was shown in Ref. \cite{BJ3} that the effective screened potential of a test charge of mass $m$ in a dense quantum plasma can be modeled using a modified Debye-H\"uckel potential also known as exponential cosine-screened Coulomb potential $V(r)=-(Z e^2/r)\exp(-r/\lambda_D)\cos(r/\lambda_D)$. It was recently shown in Ref. \cite{BJ2} that $V(r)=-(Z e^2/r)\exp(-r/\lambda_D)\cos(g\ r/\lambda_D)$ can be used to model weakly coupled plasmas with $g=0$ and dense quantum plasmas with $g=1$.

There has been a ceaseless interest (\cite{BJ4, BJ5, BJ6, BJ7} and Refs. therein) in studying atomic and molecular processes in plasmas environment due to their applications in diagnosing various plasma and also providing passable knowledge on collision dynamics. The ionization processes and atomic excitation play a crucial role in the conceptual understanding of various phenomena related to hot plasma physics, astrophysics and experiments performed with charged ions. It has been discerned that a long-range Coulomb field plays a decidedly salient role in the electron-ion scattering problem \cite{BJ4}. For instance, at small scattering angles, the total cross section for elastic scattering of a charged particle in a Coulomb field diverges; the impact excitation cross sections for electron-positive ion collisions have finite values at the reaction threshold, etc.

Strictly speaking, there are many studies focusing on studying effects of several fields on hydrogen atom embedded in plasmas.  For instance, Bahar and Soylu \cite{BJ8} studied the confinement effect of magnetic field on two-dimensional hydrogen atom in plasmas. The  plasma screening effect of dense quantum plasmas on the photodetachment cross section of hydrogen negative ion within the framework of dipole approximation was presented in \cite{BJ2}. It was found in ref. \cite{BJ10} that an anomalous resistance in plasma occurs when current flows through a plasma in a strong magnetic field. Lumb et al. \cite{BJ11} reported the effects of shape of laser pulse, confinement radius, Debye screening length as well as different laser parameters on the dynamics  of spherically confined hydrogen atom embedded in an exponential-cosine-screened Coulomb potential using the Bernstein-polynomial method. The screening and weak external electric field effects on the hydrogen atom in plasmas were also reported in ref. \cite{BJ12}. Effect of plasma screening on various properties  such as transition energy and polarizability of hydrogen like ions were studied  recently by Das \cite{BJ13}. 

Based on this information we have gathered, we have adequate materials to proceed to studying hydrogen atom in a dense and weakly coupled quantum plasmas under the influence of AB-flux field, electric field and a uniform magnetic field directed along $z$-axis. The hydrogen atom has a notable importance in quantum mechanics and quantum field theory, as a simple two-body problem which has yielded analytical solution in closed form\cite{RE1}.  Comprehension of its simple structure are very important when investigating quantum effects in more complex structures. The influences of external electric field and magnetic field on hydrogen atom has been studied in numerous papers \cite{BJ2,BJ8,BJ10,BJ11,BJ12,BJ13,RE1}. Besides using electric and magnetic fields to manipulate the energy levels or localization of quantum state of hydrogen atom in quantum plasmas, we suggest that AB-flux field could as well be used. In fact the dominance of AB-flux field on other external fields as we have shown in the manuscript, justifies its superiority. 

The Aharonov-Bohm (AB) effect is a quantum mechanical phenomenon in which an electrically charged particle is affected by an electromagnetic field despite being confined to a region in which both the magnetic field and electric field are zero. Experimental confirmation of its existence was presented in Ref. \cite{RE2}. To the best of our knowledge, there has been no previous study in which the influences of these three external fields on hydrogen atom, within a dense and weakly coupled quantum plasmas. Consequently, we feel this work will be of interest in the areas of atomic structure and collisions in plasmas.

\section{Theory and calculations}
\label{sec2}
The model equation for hydrogen atom under the influences of AB-flux, electric and uniform magnetic fields directed along $z$-axis and surrounded by quantum plasma environment can be written in cylindrical coordinates as:
\begin{widetext}
\begin{equation}
\left[\frac{1}{2\mu}\left(-i\hbar\vec{\nabla}+\frac{e}{c}\vec{A}\right)^2-\frac{Ze^2}{r}\exp\left(-\frac{r}{\lambda_D}\right)\cos\left(g\frac{r}{\lambda_D}\right)-Fr\cos(\theta)\right]\psi(r, \theta)=E_{nm}\psi(r, \theta),
\label{EE1}
\end{equation}
\end{widetext}
\begin{figure*}[!t]
 \includegraphics[height=125mm, width=190mm]{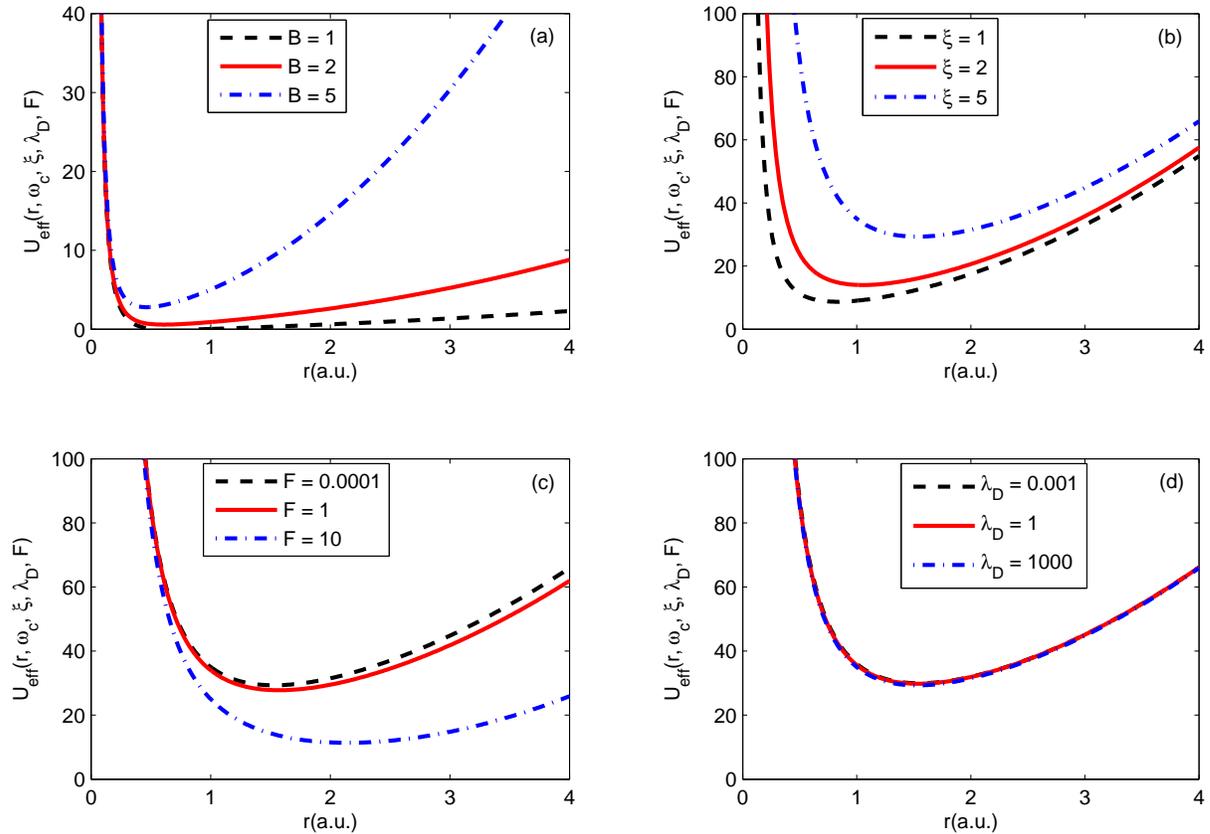}
\caption{{\protect\small (Color online) The effective potential energy to simulate dense quantum plasmas environment with rotational ($m$ = 1) levels for (a) various values of $B$ with $\xi=5, F=0.0001$ and $\lambda_D=40$. Increasing the intensity of the magnetic field  and keeping other fields constant lead to corresponding increment in the effective potential function. Thus, the potential energy becomes more repulsive. (b) Various values of $\xi$ with $B=5$, $F=0.0001$ and $\lambda_D=40$. (c) various values of $F$ with $B=5$, $\xi=5$. Increasing the strength of the electric field increase attractiveness of the effective potential. (d) Various values of $\lambda_D$ with $B=5$, $\xi=5$, $F=0.0001$. Setting effects of all fields to be constant and then vary the screening parameter up to say a factor of $1000$ have little or no effect on the effective model however it has a noticeable effect on its series expansion as it will be shown in Figure \ref{fig2}a. Moreover, suppose we neglect the effects of AB-flux field and external magnetic field as shown in Figure \ref{fig2}(b), then a significant effect of the screening parameter can be observed. This is an indication of how dominance the effects of these external fields are, on the screening parameter. All our computations are in atomic units (a.u.).}}
\label{fig1}
\end{figure*}
where $E$ denotes the energy levels, $\mu$ is the effective mass of the electron,  vector potential $\vec{A}$ may be written as a sum of two terms, $\vec{A}=\vec{A}_1+\vec{A}_2$ having the azimuthal components \cite{BJ14}: $\vec{A}_1=\frac{Br}{2}\hat{\phi}$, $\vec{A}_2=\frac{\phi_{AB}}{2\pi r}\hat{\phi}$. $\vec{B}=B\hat{z}$ is the applied external magnetic field with $\vec{\nabla}\times\vec{A}_1=\vec{B}$. $\vec{A}_2$ represents the additional magnetic flux $\phi_{AB}$ created by a solenoid inserted inside the antidot with  $\vec{\nabla}\cdot\vec{A}_2=0$.  $Z$ denotes the atomic number which is found useful in describing energy levels of light to heavy neutral atoms. In study of atomic structure, the motion of the electron in a potential created by $+Ze$ charged nuclei has been found as a very important problem. The results obtained from such studies can be applied to hydrogen atom (with $Z=1$), to He$^{+}$ (with $Z=2$) and Li$^{++}$ (with $Z=3$) \cite{BJ12}. Moreover, the characteristic properties of plasmas can be represented by the coupling parameter $\Gamma=(Ze)^2/(\alpha k_\beta T)$ (where $\alpha$ is the average distance between the particles). The ranges of electron density $n_e$ and temperature, $T$ are known as $~10^{18}-10^{23}$cm$^{-3}$ and $~10^2-10^5K$, respectively in quantum plasmas with $\Gamma>1$. Furthermore $F$ represents electric field strength with angle $\theta$ between $F$ and $r$. With $\theta=0$, then $F r\cos(\theta)$ becomes $Fr$ \cite{BJ12}. The variation of the effective potential energy as a function of various model parameters has been displayed in Figure \ref{fig1}.

Now, let us take a wavefunction in cylindrical coordinates as $\psi(r, \phi)=\frac{1}{\sqrt{2r\pi}}e^{im\phi}\mathcal{H}_{nm}(r)$, where $m=0, \pm1, \pm2, ...$ denotes the magnetic quantum number. Inserting this wavefunction into equation (\ref{EE1}), we find a second order differential equation\footnote{The detail derivation of this, for a general potential model has been shown in the Appendix A.} ${d^2\mathcal{H}_{nm}(r)}/{dr^2}+{2\mu}/{\hbar^2}\left[E_{nm}-U_{\rm eff.}\right]\mathcal{H}_{nm}(r)=0$, where the effective potential $U_{\rm eff.}$ is
\begin{widetext}
\begin{equation}
U_{\rm eff.}=-\frac{Ze^2}{r}\exp\left(-\frac{r}{\lambda_D}\right)\cos\left(g\frac{r}{\lambda_D}\right)-Fr+\frac{\omega_c\hbar}{2}(m+\xi)+\left(\frac{\mu\omega_c^2}{8}\right)r^2+\frac{\hbar^2}{2\mu}\left[\frac{(m+\xi)^2-\frac{1}{4}}{r^2}\right], 
\label{EE2}
\end{equation}
\end{widetext}
where $\xi=\phi_{AB}/\phi_{0}$ is taken as integer with the flux quantum $\phi_0=hc/e$. $\omega_c=eB/\mu c$ denotes the cyclotron frequency. In order to achieve our goal in this study, we need to solve the radial Schr\"odinger equation with the effective model (\ref{EE2}). However, the equation does not admit an exact solution with this model. Thus, we are constrain to utilize two methods: numerical procedure or perturbation technique. In this paper, we employ a perturbative formalism \cite{BJ15} to solve the problem. In this perturbation technique, it is mandatory to split the effective model into two sub models. The main part should corresponds to a shape invariant potential in which the superpotential is known analytically and the second part will be considered as the perturbation. This approach has been employed by Ikhdair and Sever to obtain bound state energy for the exponential-cosine-screened Coulomb potential \cite{BJ16}. Now, in view of the above information, let us re-write the wave function $\mathcal{H}_{nm}(r)$ to reflect the known normalized eigenfunction of the unperturbed system ($\mathcal{P}_{nm}(r)$) and moderating function ($\mathcal{Q}_{nm}(r)$) corresponding to the perturbation potential, in the form $\mathcal{H}_{nm}(r)=\mathcal{P}_{nm}(r)\mathcal{Q}_{nm}(r)$. Substitution of this expression into the radial Schr\"{o}dinger equation gives
\begin{widetext}
\begin{equation}
\frac{\hbar^2}{2\mu}\left[\frac{1}{\mathcal{P}_{nm}(r)}\frac{d^2\mathcal{P}_{nm}(r)}{dr^2}+\frac{1}{\mathcal{Q}_{nm}(r)}\frac{d^2\mathcal{Q}_{nm}(r)}{dr^2}+\frac{2}{\mathcal{P}_{nm}(r)\mathcal{Q}_{nm}(r)}\frac{d\mathcal{P}_{nm}(r)}{dr}\frac{d\mathcal{Q}_{nm}(r)}{dr}\right]=U_{\rm eff.}(r)-E_{nm}.
\label{EE3}
\end{equation}
The logarithmic derivatives of the perturbed and unperturbed wave functions can be written as follow:
\begin{equation}
\Delta W_{nm} = -\frac{\hbar}{\sqrt{2\mu}}\frac{1}{\mathcal{Q}_{nm}(r)}\frac{d\mathcal{Q}_{nm}(r)}{dr}\ \ \ \mbox{and}\ \ \ W_{nm}=-\frac{\hbar}{\sqrt{2\mu}}\frac{1}{\mathcal{P}_{nm}(r)}\frac{d\mathcal{P}_{nm}(r)}{dr},
\label{EE4}
\end{equation}
\begin{figure*}[!t]
 \includegraphics[height=125mm, width=190mm]{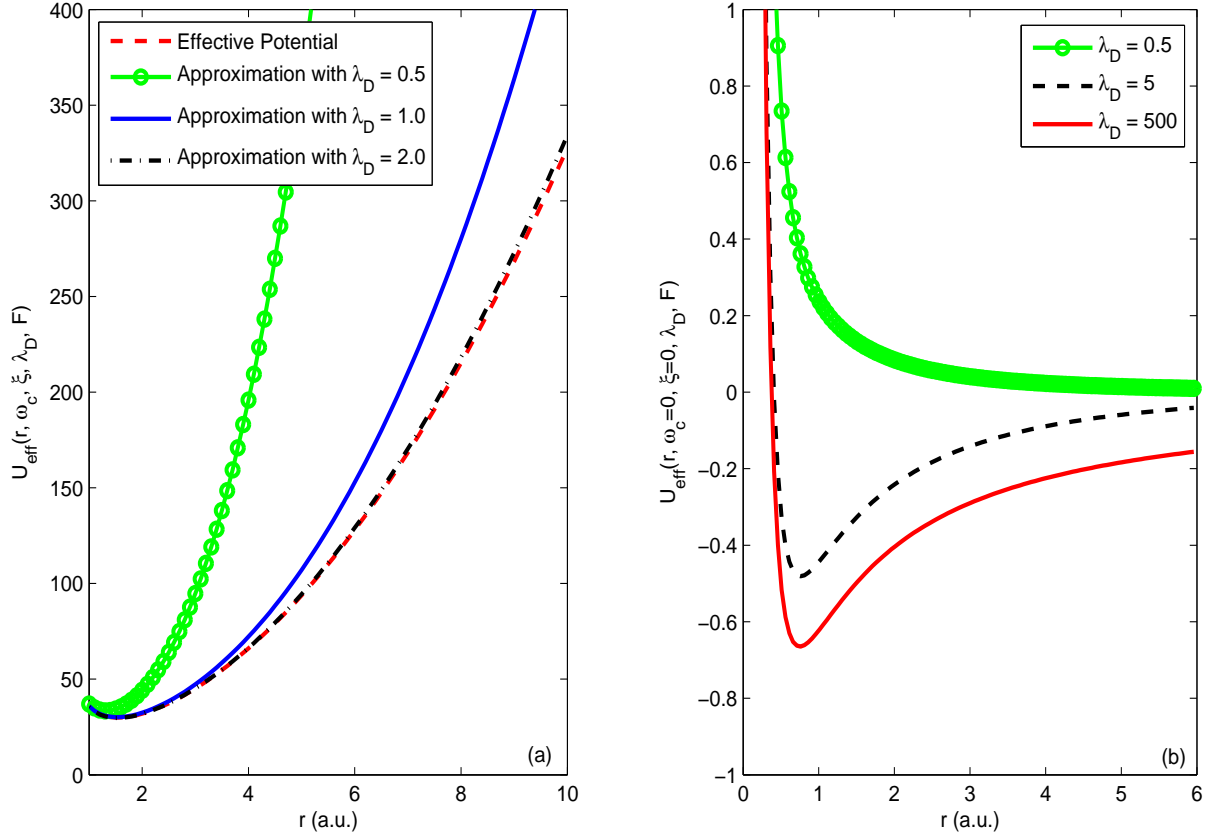}
\caption{{\protect\small (Color online) (a) Variation of the effective potential and its approximation with $r$. We have used the following fitting parameters $g=1$, $\xi=5$ and $F=0.001$. The approximation is only valid for small $\lambda_D^{-1}$. For this reason, we shall choose $\lambda_D\geq2$ where necessary in our computation. However, one may wonder why the series have been truncated at order 3. With respect to this, convergence is not an important property for series approximations in physical problems. A slowly convergent approximation which requires many terms to achieve reasonable accuracy is much less valuable than the divergent series which gives accurate answers in a few terms \cite{BJ16}. This is the reason why we truncate the series expansion in equation (\ref{EE9}) at a lower order term. (b) The effective potential energy to simulate dense quantum plasmas environment with rotational ($m = 1$) levels for various values of $\lambda_D$ with $F = 0.0001$. All our computations are in a.u.}}
\label{fig2}
\end{figure*}
respectively, which result to
\begin{equation}
\frac{\hbar^2}{2\mu\mathcal{P}_{nm}(r)}\frac{d^2\mathcal{P}_{nm}(r)}{dr^2}=W_{nm}^2-\frac{\hbar}{\sqrt{2\mu}}\frac{dW_{nm}}{dr}=\left(V_0(r)+\frac{\hbar^2}{2\mu}\left[\frac{\left(\sigma_{0m}-\frac{1}{2}\right)^2-\frac{1}{4}}{r^2}\right]\right)-\epsilon_{nm}
\label{EE5}
\end{equation}
and
\begin{eqnarray}
\frac{\hbar^2}{2\mu}\left[\frac{1}{\mathcal{Q}_{nm}(r)}\frac{d^2\mathcal{Q}_{nm}(r)}{dr^2}+\frac{2}{\mathcal{P}_{nm}(r)\mathcal{Q}_{nm}(r)}\frac{d\mathcal{P}_{nm}(r)}{dr}\frac{d\mathcal{Q}_{nm}(r)}{dr}\right]=\Delta W_{nm}^2-\frac{\hbar}{\sqrt{2\mu}}\frac{d\Delta W_{nm}}{dr}\label{EE6}
+2W_{nm}\Delta W_{nm}=\Delta U_{\rm eff.}(r)-\Delta\epsilon_{nm}, \nonumber\\
\end{eqnarray}
\end{widetext}
where $V_0(r)$ denotes the unperturbed potential, $\epsilon_{nm}$ represents the eigenvalues of the unperturbed system and $\Delta\epsilon_{nm} = E_{nm}^{(1)}+E_{nm}^{(2)}+E_{nm}^{(3)}+...$ is the energy eigenvalues of the perturbed system which provide correction term to the energy such that the total eigenvalues become $E_{nm}=\epsilon_{nm}+\Delta\epsilon_{nm}$. By comparing supersymmetric perturbation theory with the logarithmic perturbation theory \cite{BJ17}, equation (\ref{EE6}) seems to be in a closed analytical form. This made the approach utilized in this study to be more advantageous than those available in the literature \cite{BJ17, BJ18, BJ19}.

As we have mentioned earlier, it is necessary to split the effective potential into two parts. Within this context, the zeroth order term corresponds to the Coulomb potential while the higher order terms constitute the perturbation expressions. However, the perturbation equation (\ref{EE5}) cannot be exactly solved in its present form. It is therefore required to expand the related functions to the perturbation in terms of the perturbation parameter {$\eta$} (which we shall eventually set as unity):
\begin{eqnarray}
&&\Delta U_{\rm eff.}(r;\eta)=\sum_{i=1}^\infty\eta_iU_{\rm eff.}(r)^{(i)}, \ \ \Delta W_{nm}(r;\eta)=\sum_{i=1}^\infty\eta_iW_{nm}^{(i)},\nonumber\\
&&\Delta E_{nm}^{(i)}(\eta)=\sum_{i=1}^\infty\eta_iE_{nm}^{(i)},
\label{EE7}
\end{eqnarray}
where $i$ represents the order of perturbation. We substitute equation (\ref{EE7}) into equation (\ref{EE6}) and then equate terms with same power of $\eta$ on both sides to have the following expressions
\begin{widetext}
\begin{subequations}
\begin{eqnarray}
2W_{nm}(r)W_{nm}^{(1)}(r)-\frac{\hbar}{\sqrt{2\mu}}\frac{dW_{nm}^{(1)}(r)}{dr}=V_1(r)-E_{nm}^{(1)}, \label{EE8a}\\
W_{nm}^{(1)2}(r)+2W_{nm}(r)W_{nm}^{(2)}(r)-\frac{\hbar}{\sqrt{2\mu}}\frac{dW_{nm}^{(2)}(r)}{dr}=V_2(r)-E_{nm}^{(2)}, \label{EE8b}\\
2\left[W_{nm}(r)W_{nm}^{(3)}(r)+W_{nm}^{(1)}(r)W_{nm}^{(2)}(r)\right]-\frac{\hbar}{\sqrt{2\mu}}\frac{dW_{nm}^{(3)}(r)}{dr}=V_3(r)-E_{nm}^{(3)}, \label{EE8c}\\
2\left[W_{nm}(r)W_{nm}^{(4)}(r)+W_{nm}^{(1)}(r)W_{nm}^{(3)}(r)\right]+W_{nm}^{(2)}(r)W_{nm}^{(2)}(r)-\frac{\hbar}{\sqrt{2\mu}}\frac{dW_{nm}^{(4)}(r)}{dr}=V_4(r)-E_{nm}^{(4)}.
\label{EE8d}
\end{eqnarray}
\end{subequations}
\end{widetext}
It is now time to apply this background information to our problem. Thus, the effective potential in equation (\ref{EE2}) can be expanded in power series of the Debye screening parameter $\lambda_D$ as
\begin{widetext}
\begin{eqnarray}
U_{\rm eff.}(r)&=&-\frac{A}{r}+\frac{\hbar^2}{2\mu}\left[\frac{(m+\xi)^2-\frac{1}{4}}{r^2}\right]+\left[\frac{A}{\lambda_D}+\frac{\omega_c\hbar}{2}(m+\xi)\right]-\left[F+\frac{A}{\lambda_D^2}\left(\frac{1}{2}-\frac{g^2}{2}\right)\right]r\nonumber\\
&&+\left[\frac{A}{\lambda_D^3}\left(\frac{1}{6}-\frac{g^2}{2}\right)+\frac{\mu\omega_c^2}{8}\right]r^2-\left[\frac{A}{\lambda_D^4}\left(\frac{1}{24}-\frac{g^2}{4}+\frac{g^4}{24}\right)\right]r^3+\mathcal{O}(r^4),
\label{EE9}
\end{eqnarray}
\end{widetext}
where $A=Ze^2$. The accuracy of this approximation has been shown in Figure \ref{fig2}a.  It is only valid for large value of Debye screening parameter $\lambda_D$. The first term in the series expansion (\ref{EE9}) is the unperturbed term which is Coulomb potential with a well known solution. Alternatively, it can be easily obtained using the recently proposed formula method \cite{BJ20}. The second term is centrifugal term while the remaining terms constitute the perturbation expression. Within this context, the unperturbed energy and the corresponding normalized wave function can be written as:
\begin{eqnarray}
E_{nm}^{(0)}&=&-\frac{\mu}{2\hbar^2}\frac{A^2}{\sigma_{nm}^2}\ \ \mbox{and}\nonumber\\ \mathcal{P}_{nm}^{(0)}(r)&=&N_{nm}^{(0)}r^{\sigma_{0m}}e^{-\varrho r}\ L_{n}^{2\sigma_{0m}-1}\left[2\varrho r\right],
\label{EE10}
\end{eqnarray}
respectively with $n = 0, 1, 2, ...$. The ground state superpotential and the normalization factor can be written as
\begin{eqnarray}
&&W_{n=0, m}^{(0)}(r)=-\frac{\hbar}{\sqrt{2\mu}}\frac{\sigma_{0m}}{r}+\frac{1}{\hbar}\sqrt{\frac{\mu}{2}}\frac{A}{\sigma_{0m}},\nonumber \\ 
&&N_{nm}^{(0)}=\frac{\left(2\varrho \right)^{\sigma_{0m}}}{\sigma_{nm}}\left[\frac{\hbar^2\left(2\sigma_{nm}-1-n\right)!}{\mu n!A }\right]^{-\frac{1}{2}},
\label{EE11}
\end{eqnarray}
where $\varrho={A\mu}/(\hbar^2\sigma_{nm})$. Now, let us consider the expressions leading to the first-, second- and third-order perturbations given by equations (\ref{EE8a}-\ref{EE8d}). Using superpotentials given in equation (\ref{EE4}) and multiplying each term in equations (\ref{EE8a}-\ref{EE8d}) by ${P_{nm}^{(0)}}^2$, we obtain first-order correction to the energy and its superpotential as follows:
\begin{widetext}
\begin{eqnarray}
E_{nm}^{(1)}&=&\int_{-\infty}^{\infty}{\mathcal{P}_{nm}^{(0)}}^2(r)\left(-\left[F+\frac{A}{\lambda_D^2}\left(\frac{1}{2}-\frac{g^2}{2}\right)\right]r\right)dr,
\label{EE12a}\\
W_{nm}^{(1)}&=&\sqrt{\frac{2\mu}{\hbar}}\frac{1}{{\mathcal{P}_{nm}^{(0)}}^2(r)}\int_{}^{r}{\mathcal{P}_{nm}^{(0)}}^2(y)\left(E_{nm}^{(1)}+\left[F+\frac{A}{\lambda_D^2}\left(\frac{1}{2}-\frac{g^2}{2}\right)\right]y\right)dy.
\label{EE12b}
\end{eqnarray}
Also, the second-order correction to the energy and its superpotential can be written as follows:
\begin{subequations}
\begin{eqnarray}
E_{nm}^{(2)}&=&\int_{-\infty}^{\infty}{\mathcal{P}_{nm}^{(0)}}^2(r)\left(\left[\frac{A}{\lambda_D^3}\left(\frac{1}{6}-\frac{g^2}{2}\right)+\frac{\mu\omega_c^2}{8}\right]r^2-{W_{nm}^{(1)}}^2\right)dr,
\label{EE13a}\\
W_{nm}^{(2)}&=&\sqrt{\frac{2\mu}{\hbar}}\frac{1}{{\mathcal{P}_{nm}^{(0)}}^2(r)}\int_{}^{r}{\mathcal{P}_{nm}^{(0)}}^2(y)\left(E_{nm}^{(2)}-\left[\frac{A}{\lambda_D^3}\left(\frac{1}{6}-\frac{g^2}{2}\right)+\frac{\mu\omega_c^2}{8}\right]y^2+{W_{nm}^{(1)}}^2\right)dy.\nonumber\\
\label{EE13b}
\end{eqnarray}
\end{subequations}
The third-order correction to the energy and its superpotential become:
\begin{subequations}
\begin{eqnarray}
E_{nm}^{(3)}&=&\int_{-\infty}^{\infty}{\mathcal{P}_{nm}^{(0)}}^2(r)\left(\left[\frac{A}{\lambda_D^4}\left(\frac{1}{24}-\frac{g^2}{4}+\frac{g^4}{24}\right)\right]r^3+W_{nm}^{(1)}(r)W_{nm}^{(2)}(r)\right)dr, \label{EE14a}\\
W_{nm}^{(2)}(r)&=&\sqrt{\frac{2\mu}{\hbar^2}}\frac{1}{{\mathcal{P}_{nm}^{(0)}}^2(r)}\int_{}^{r}{\mathcal{P}_{nm}^{(0)}}^2(y)\left(E_{nm}^{(3)}+{W_{nm}^{(1)}(r)}{W_{nm}^{(2)}(r)}+\left[\frac{A}{\lambda_D^4}\left(\frac{1}{24}-\frac{g^2}{4}+\frac{g^4}{24}\right)\right]y^3\right)dy.\nonumber\\ \label{EE14b}
\end{eqnarray}
\end{subequations}
\end{widetext}
Using the expressions for $E_{nm}^{(1)}$, $E_{nm}^{(2)}$ and $E_{nm}^{(3)}$, one can calculate superpotentials $W_{nm}^{(1)}$, $W_{nm}^{(2)}$ and $W_{nm}^{(3)}$ explicitly. Consequently, the superpotentials  can be used to calculate the moderating wave function $\mathcal{Q}_{nm}(r)\approx\exp\left(-{\sqrt{2\mu}}/{\hbar}\int^r\left(W_{nm}^{(1)}(r)+W_{nm}^{(2)}(r)\right)\right)$. Since we now have all necessary formulas needed for our calculation, let us now focus our attention on how to utilize them to deduce ground and exited state energies together with their moderating superpotentials. We start with the ground state, such that from equations (\ref{EE12a}-\ref{EE14b}), we have:
\begin{widetext}
\begin{eqnarray}
&&E_{0m}^{(1)}=-\frac{\hbar^2}{2\mu}\left\{3\sigma_{0m}^2-\left(\sigma_{0m}-\frac{1}{2}\right)^2+\frac{1}{4}\right\}\left\{\frac{F}{A}+\frac{1}{\lambda_D^2}\left(\frac{1}{2}-\frac{g^2}{2}\right)\right\}\ \ \mbox{and}\ \ \ W_{0m}^{(1)}(r)=-\frac{\hbar}{\sqrt{2\mu}}\left\{\frac{F}{A}+\frac{1}{\lambda_D^2}\left(\frac{1}{2}-\frac{g^2}{2}\right)\right\}\sigma_{0m}r.\nonumber\label{EE15a}\\
&&E_{0m}^{(2)}=\left\{\left[\frac{A}{\lambda_D^3}\left(\frac{1}{6}-\frac{g^2}{2}\right)+\frac{\mu\omega_c^2}{8}\right]-\frac{\hbar^2\sigma_{0m}^2}{2\mu}\left[\frac{F}{A}+\frac{1}{\lambda_D^2}\left(\frac{1}{2}-\frac{g^2}{2}\right)\right]^2\right\}\times\frac{\hbar^4\sigma_{0m}^2}{{2\mu^2A^2}}\left\{5\sigma_{0m}^2-3\left(\sigma_{0m}-\frac{1}{2}\right)^2+\frac{7}{4}\right\}\ \ \mbox{and}\nonumber\\
&&W_{0m}^{(2)}(r)=-\left\{\frac{\hbar^2\sigma_{0m}^2}{{2\mu}}\left[\frac{F}{A}+\frac{1}{\lambda_D^2}\left(\frac{1}{2}-\frac{g^2}{2}\right)\right]^2-\left(\frac{A}{\lambda_D^3}\left(\frac{1}{6}-\frac{g^2}{2}\right)+\frac{\mu\omega_c^2}{8}\right)\right\}\times\frac{\hbar r\sigma_{0m}}{\mu A^2\sqrt{2\mu}}\left[\sigma_{0m}\sigma_{1}\hbar^2+\mu Ar\right].\label{EE15b}
\end{eqnarray}
\end{widetext}
Consequently, the approximate expressions for the ground state energy and radial wavefunction of hydrogen atom in AB-flux, electric and uniform magnetic fields directed along $z$-axis and surrounded by quantum plasmas, can be written as
\begin{widetext}
\begin{eqnarray}
E_{0m}\approx E_{0m}^{(0)}+\left[\frac{A}{\lambda_D}+\frac{\omega_c\hbar}{2}(m+\xi)\right]+E_{0m}^{(1)}+E_{0m}^{(2)}\label{E16a}\ \ \ \ \mbox{and}\ \ 
\psi(r, \phi)\approx\frac{1}{\sqrt{2r\pi}}e^{im\phi}\mathcal{P}_{nm}(r)\exp\left(-\frac{\sqrt{2\mu}}{\hbar}\int^r\left(W_{0m}^{(1)}+W_{0m}^{(2)}\right)\right).\nonumber\\
\label{E16b}
\end{eqnarray}
\end{widetext}
It is worth mentioning that, there is a corresponding relationship between 2D and 3D which can be obtained by making a replacement $m+\xi = \ell + 1/2$. Therefore, the bound state energy levels for $1$s in the absent of external magnetic and AB-flux field in 3D can also be deduced from the above equations.

Let us now proceed to excited state calculations. We calculate the energy shift and superpotentials of first and second -order as follows:
\begin{widetext}
\begin{eqnarray}
&&E_{1m}^{(1)}=-\frac{\hbar^2}{2\mu}\left\{3\sigma_{1m}^2-\left(\sigma_{0m}-\frac{1}{2}\right)^2+\frac{1}{4}\right\}\left\{\frac{F}{A}+\frac{1}{\lambda_D^2}\left(\frac{1}{2}-\frac{g^2}{2}\right)\right\}\ \ \mbox{and}\ \ \ W_{1m}^{(1)}(r)=-\frac{\hbar}{\sqrt{2\mu}}\left\{\frac{F}{A}+\frac{1}{\lambda_D^2}\left(\frac{1}{2}-\frac{g^2}{2}\right)\right\}\sigma_{1m}r.\nonumber\label{EE17a}\\
&&E_{1m}^{(2)}=\left\{\left[\frac{A}{\lambda_D^3}\left(\frac{1}{6}-\frac{g^2}{2}\right)+\frac{\mu\omega_c^2}{8}\right]-\frac{\hbar^2\sigma_{1m}^2}{2\mu}\left[\frac{F}{A}+\frac{1}{\lambda_D^2}\left(\frac{1}{2}-\frac{g^2}{2}\right)\right]^2\right\}\times\frac{\hbar^4\sigma_{1m}^2}{{2\mu^2A^2}}\left\{5\sigma_{1m}^2-3\left(\sigma_{0m}-\frac{1}{2}\right)^2+\frac{7}{4}\right\}\ \ \mbox{and}\nonumber\\
&&W_{1m}^{(2)}(r)=-\left\{\frac{\hbar^2\sigma_{1m}^2}{{2\mu}}\left[\frac{F}{A}+\frac{1}{\lambda_D^2}\left(\frac{1}{2}-\frac{g^2}{2}\right)\right]^2-\left(\frac{A}{\lambda_D^3}\left(\frac{1}{6}-\frac{g^2}{2}\right)+\frac{\mu\omega_c^2}{8}\right)\right\}\times\frac{\hbar r\sigma_{1m}}{\mu A^2\sqrt{2\mu}}\left[\sigma_{1m}\sigma_{2m}\hbar^2+\mu Ar\right].\label{EE17b}
\end{eqnarray}
\end{widetext}
Therefore, the approximated energy eigenvalues of the hydrogen atom in AB-flux, electric and uniform magnetic fields directed along $z$-axis and surrounded by quantum plasmas, corresponding to the first excited state ($n = 1$) are:
\begin{equation}
E_{1m}\approx E_{1m}^{(0)}+\left[\frac{A}{\lambda_D}+\frac{\omega_c\hbar}{2}(m+\xi)\right]+E_{1m}^{(1)}+E_{1m}^{(2)}.\label{EE18}
\end{equation}
We can proceed further to obtain expression for states $n=2, 3, 4, 5, \ldots.$ However, we leave the calculations as exercises in elementary integrals. From the supersymmetry, we can write out the $n$-th state energy shifts and the corresponding superpotentials as:
\begin{widetext}
\begin{eqnarray}
&&E_{nm}^{(1)}=-\frac{\hbar^2}{2\mu}\left\{3\sigma_{nm}^2-\left(\sigma_{0m}-\frac{1}{2}\right)^2+\frac{1}{4}\right\}\left\{\frac{F}{A}+\frac{1}{\lambda_D^2}\left(\frac{1}{2}-\frac{g^2}{2}\right)\right\}\ \ \mbox{and}\ \ \ W_{nm}^{(1)}(r)=-\frac{\hbar}{\sqrt{2\mu}}\left\{\frac{F}{A}+\frac{1}{\lambda_D^2}\left(\frac{1}{2}-\frac{g^2}{2}\right)\right\}\sigma_{nm}r.\nonumber\label{EE19a}\\
&&E_{nm}^{(2)}=\left\{\left[\frac{A}{\lambda_D^3}\left(\frac{1}{6}-\frac{g^2}{2}\right)+\frac{\mu\omega_c^2}{8}\right]-\frac{\hbar^2\sigma_{nm}^2}{2\mu}\left[\frac{F}{A}+\frac{1}{\lambda_D^2}\left(\frac{1}{2}-\frac{g^2}{2}\right)\right]^2\right\} \times\frac{\hbar^4\sigma_{nm}^2}{{2\mu^2A^2}}\left\{5\sigma_{nm}^2-3\left(\sigma_{0m}-\frac{1}{2}\right)^2+\frac{7}{4}\right\}\ \ \mbox{and}\nonumber\\
&&W_{nm}^{(2)}(r)=-\left\{\frac{\hbar^2\sigma_{nm}^2}{{2\mu}}\left[\frac{F}{A}+\frac{1}{\lambda_D^2}\left(\frac{1}{2}-\frac{g^2}{2}\right)\right]^2-\left(\frac{A}{\lambda_D^3}\left(\frac{1}{6}-\frac{g^2}{2}\right)+\frac{\mu\omega_c^2}{8}\right)\right\}\times\frac{\hbar r\sigma_{nm}}{\mu A^2\sqrt{2\mu}}\left[\sigma_{nm}\sigma_{n+1, m}\hbar^2+\mu Ar\right].\label{EE19b}
\end{eqnarray}
\end{widetext}
Consequently, we obtain the approximate energy eigenvalues of the hydrogen atom in quantum plasmas environment under the influences of AB-flux, electric and uniform magnetic fields, directed along $z$-axis, corresponding to the $n$th-state as
\begin{equation}
E_{nm}\approx E_{nm}^{(0)}+\left[\frac{A}{\lambda_D}+\frac{\omega_c\hbar}{2}(m+\xi)\right]+E_{nm}^{(1)}+E_{nm}^{(2)}.\label{EE20}
\end{equation}
\begin{table*}[!t]
{\scriptsize
\caption{\footnotesize The energy values for hydrogen atom in dense quantum plasma under the influence of AB-flux, external magnetic and electric fields with various values of magnetic quantum numbers. The following fitting parameters have been employed:  $A=1$, $\lambda_D=20$, $m=1$ and $g=1$. Our computations are performed with respect to a.u.}\vspace*{10pt}{
\begin{ruledtabular}
\begin{tabular}{ccccccc}
{}&{}&{}&{}&{}&{}&{}\\[-1.0ex]
m	&n	&$F=0$, $\xi=0$, $B=0$	&$F=0$, $\xi=0$, $B=5$	&$F=0$, $\xi=5$, $B=0$	&$F=5$, $\xi=0$, $B=0$	&$F=5$, $\xi=5$, $B=5$\\[1ex]\hline
0	&0	&-1.95001560	&-0.7781406	&-0.0156852	&-5.6218906	&-442558.77	\\[1ex]
	&1	&-0.17283160	&45.530293	&-0.0833031	&-429.00096	&-1530748.8	\\[1ex]
	&2	&-0.03429688	&322.23133	&-0.2026389	&-8104.1749	&-4095347.7	\\[1ex]
	&3	&-0.00689445	&1205.8525	&-0.3904204	&-59179.616	&-9364067.9	\\[1ex]
	{}&{}&{}&{}&{}&{}&{}						\\[1ex]
1	&0	&-0.17269097	&37.483559	&-0.0542562	&-331.57894	&-1164728.2	\\[1ex]
	&1	&-0.03390625	&295.43484	&-0.1639670	&-7369.2527	&-3445630.2	\\[1ex]
	&2	&-0.00612883	&1150.9314	&-0.3407485	&-56363.444	&-8291125.3	\\[1ex]
	&3	&-0.01687886	&3166.5456	&-0.6042120	&-256439.08	&-17514714	\\[1ex]
	{}&{}&{}&{}&{}&{}&{}						\\[1ex]
-1&0	&-1.95000000	&-4.4500000	& 0.0021055	&-1.9500000	&-139321.25	\\[1ex]
	&1	&-1.95000000	&-4.4500000	&-0.0327008	&-1.9500000	&-595724.41	\\[1ex]
	&2	&-0.17269097	&32.483559	&-0.1070687	&-331.57894	&-1830220.7	\\[1ex]
	&3	&-0.03390625	&290.43484	&-0.2342795	&-7369.2527	&-4626935.2	\\[1ex]
\end{tabular}\label{tab1}
\end{ruledtabular}}
\vspace*{-1pt}}
\end{table*}
\begin{table*}[!t]
{\scriptsize
\caption{\footnotesize The energy values for hydrogen atom in a weakly coupled plasmas under the influence of AB-flux, external magnetic and electric fields with various values of magnetic quantum numbers. The following fitting parameters have been employed: $A=1$, $\lambda_D=20$, $m=1$ and $g=0$. Our computations are performed with respect to a.u.} \vspace*{10pt}{
\begin{ruledtabular}
\begin{tabular}{ccccccc}
{}&{}&{}&{}&{}&{}&{}\\[-1.0ex]
m	&n	&$F=0$, $\xi=0$, $B=0$	&$F=0$, $\xi=0$, $B=5$	&$F=0$, $\xi=5$, $B=0$	&$F=5$, $\xi=0$, $B=0$	&$F=5$, $\xi=5$, $B=5$\\[1ex]\hline
0	&0	&-1.9506173	&-0.7787423	&-0.0110816	&-5.6230782	&-442781.81	\\[1ex]
	&1	&-0.1763182	&45.526807	&-0.0610759	&-429.21011	&-1531518.6	\\[1ex]
	&2	&-0.0402301	&322.22539	&-0.1840939	&-8108.2092	&-4097404.3	\\[1ex]
	&3	&-0.0095952	&1205.8498	&-0.4474275	&-59209.163	&-9368766.0	\\[1ex]
	{}&{}&{}&{}&{}&{}&{}						\\[1ex]
1	&0	&-0.1757576	&37.480492	&-0.0457138	&-331.74021	&-1165313.9	\\[1ex]
	&1	&-0.0397546	&295.42900	&-0.1557682	&-7372.9206	&-3447360.5	\\[1ex]
	&2	&-0.0091772	&1150.9283	&-0.3980985	&-56391.583	&-8295285.1	\\[1ex]
	&3	&-0.0071157	&3166.5554	&-0.8757537	&-256567.21	&-17523495	\\[1ex]
	{}&{}&{}&{}&{}&{}&{}						\\[1ex]
-1&0	&-1.9500000	&-4.4500000	&-0.0000247	&-1.9500000	&-139391.73	\\[1ex]
	&1	&-1.9500000	&-4.4500000	&-0.0178498	&-1.9500000	&-596024.65	\\[1ex]
	&2	&-0.1757576	&32.480492	&-0.0736449	&-331.74021	&-1831141.1	\\[1ex]
	&3	&-0.0397546	&290.42900	&-0.2072696	&-7372.9206	&-4629258.8	\\[1ex]
\end{tabular}\label{tab2}
\end{ruledtabular}}
\vspace*{-1pt}}
\end{table*}
\begin{figure*}[!t]
\includegraphics[height=85mm, width=180mm]{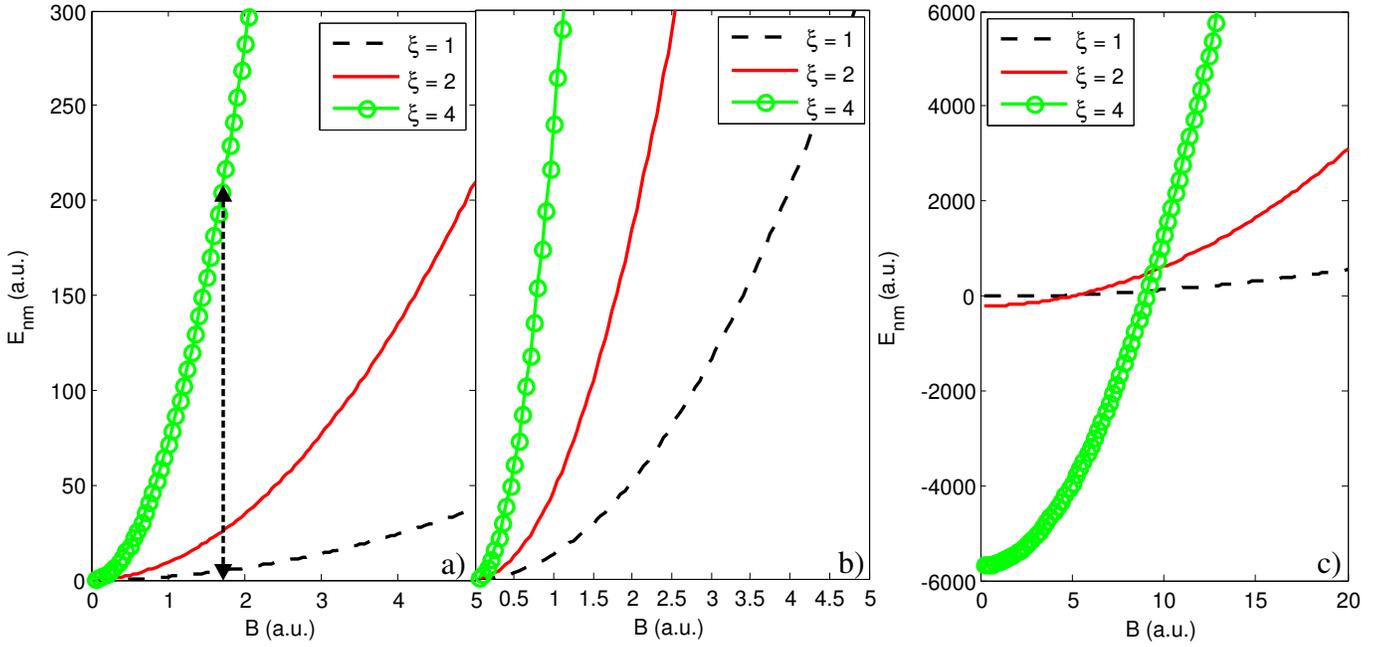}
\caption{{\protect\footnotesize Variation of energy values for hydrogen atom in quantum plasmas and under the influence of magnetic field, AB flux field and electric field in atomic units using the fitting parameters $m = n =0$ and $\lambda_D=20$ (a) as a function of external magnetic field with various $xi$ and with $F=0.0001$. (b) Same as (a) but with $m=-1$ and $n=2$. (c) same as (a) but with $F=1.2$. All our computations are expressed with respect to a.u.}}
\label{fig3}
\end{figure*}
\begin{figure*}[!t]
\includegraphics[height=85mm, width=180mm]{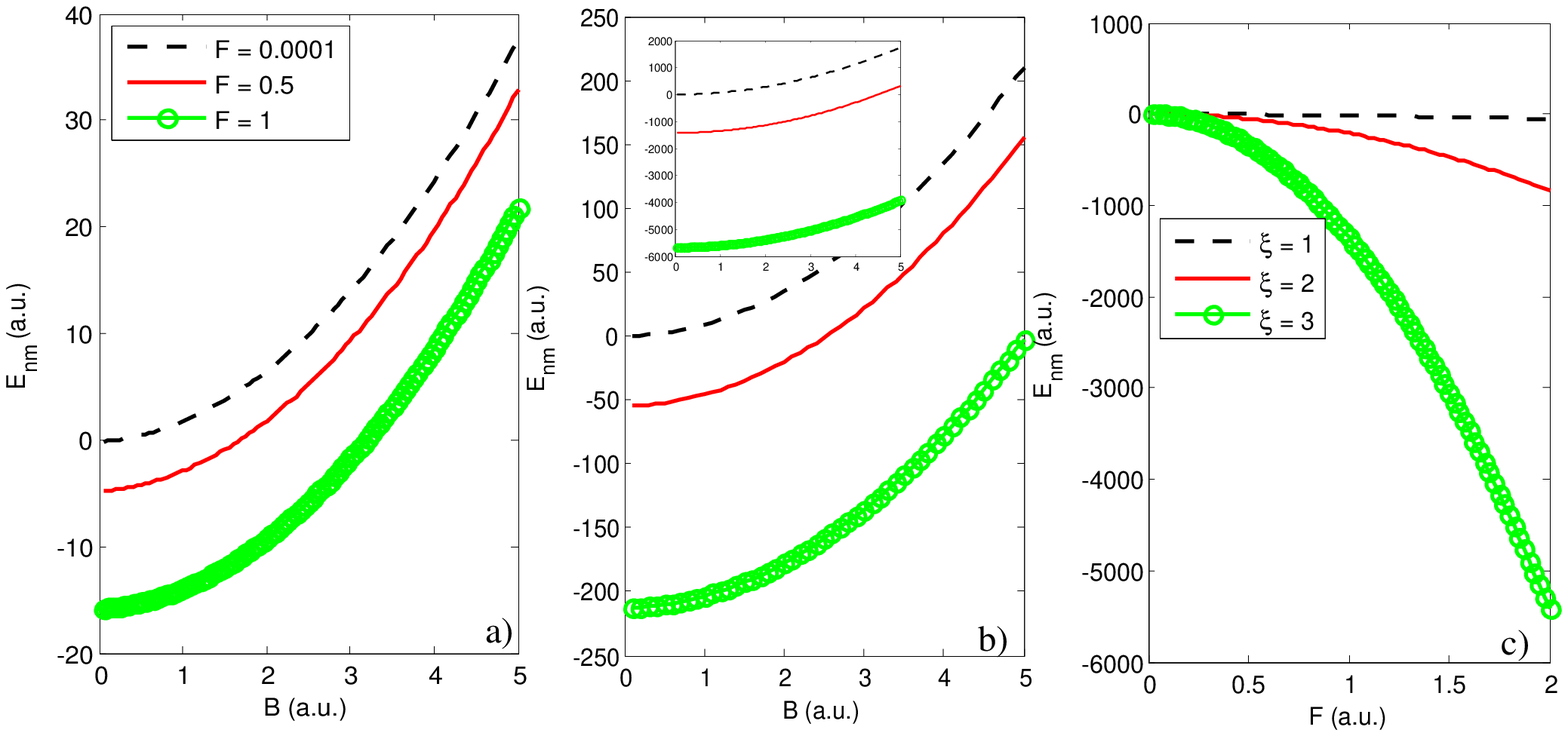}
\caption{{\protect\footnotesize Variation of energy values for hydrogen atom in quantum plasmas and under the influence of magnetic field, AB flux field and electric field in atomic units using the fitting parameters $m = n =0$ and $\lambda_D=20$ (a) as a function of external magnetic field with various $F$ and with $\xi=1$ (b) Same as (a) but with $\xi=2$ while the sub figure is for $\xi=4$. (c) as a function of magnitude of external electric field with various $\xi$ and with $B=1$}}
\label{fig4}
\end{figure*}
It is worth mentioning that in all our calculations for energy, we have changed the lower limit of the integration from $-\infty$ to $0$ so as to accommodate the fact that $r$ is never negative. Tables \ref{tab1} and \ref{tab2} display eigenvalues for hydrogen atom in quantum plasmas under the influence and the absence of external fields (magnetic field, AB flux field and electric field) in a.u. and in low vibrational $n$ and rotational $m$. From the tables, in the absence of external fields, (i.e., when $B = \xi = F = 0$), the spacing between the energy levels of the effective potential is narrow and decrease with increasing $n$. We notice that, there exists degeneracy among some states (n, m). For instance, $(1, 1)$ and $(3, -1)$; $(0, 1)$ and $(2, -1)$ also quasi-degeneracy of the energy levels among some states. For instance $(2, 0)$ and $(1, 1)$; $(2, 1)$ and $(3, 0)$;  but application of magnetic field strength does not only increase the energy levels of the effective potential and spacings between states but also transform the degeneracies to quasi-degeneracy. Moreover, the quasi-degeneracies among the states are also removed and the energy values shift up.

By subjecting the hydrogen atom in quantum plasmas to only AB flux field, reduced the energy values and degeneracies are removed, whereas the quasi-degeneracies among the states are not affected. The energy levels become more negative and the system becomes strongly attractive as quantum number $n$ increase for fixed $m$. When only electric field is applied, the degeneracies and the quasi-degeneracies are not affected and the attractiveness of the total interaction potential increase. The overall effects indicate that the system is strongly attractive while the localizations of quantum levels change and the eigenvalues decrease.  Also, the combined effect of the fields is stronger than the individual effects and consequently, there is a considerably shift in the bound state energy of the system.

In Figure \ref{fig3}, we have studied the combined effect of AB-flux, magnetic and electric fields on the energy values of hydrogen atom in quantum plasmas environment. The confinement effect of AB flux field on the hydrogen atom in quantum plasmas is stronger than that of magnetic field. This can be seen in (a) by comparing the energy values when $B$ is $1.7$ and when $\xi$ is $1$, $2$ and $4$. For instance, when AB flux field is small, say $\xi=1$ and then increase the intensity of the magnetic field, it can be seen that energy shift is $\approx5$. However, within same range of magnetic field intensity, a little distortion, say $\xi=2$ leads to a huge shift in energy level. Figure (b) also shows similar properties. However in (c), we study the effect of high electric field intensity on the hydrogen atom in quantum plasmas. As it can be seen, a low intensity of AB-flux field, say $\xi=1$ and high electric field ($F=1.2$) can not affect hydrogen atom in quantum plasmas, even while increasing the magnetic field intensity gradually. However, suppose we adjust the intensity of AB-flux field (say $\xi=3$), a very low energy can be obtained provided that the magnetic field is low. However, as the intensity of the magnetic field increase so does the energy.  This indicates that the energy values of hydrogen atom in quantum plasmas environment or localization of quantum states can be changed or adjusted to a maximum level by applying a strong magnetic field and electric field intensity. A weak magnetic field and strong electric field intensity will reduce the energy of the hydrogen atom to a minimum level. In either way, AB-flux field can act as catalyst to boost the process.

Figure {\ref{fig4}} displays the dominance of AB-flux field on external electric and magnetic fields. This can be seen via comparison of figures (a), (b) and (c). From (a), it can be seen that when hydrogen atom is under a low AB-flux field, the gap between energy levels when $F=0.0001$ and $1.2$ is tiny. However in (b), where we increase the intensity a little (say, $\xi=2$), it can be observed that the energy values of the process gradually increase from negative to positive as magnetic field increase. Moreover, the gap between $F=0.0001$ and $1.2$ becomes wide. Furthermore, we double the intensity of the formal AB-flux (i.e., $\xi=4$) as displayed in subplot of (b). We observe that for a very weak electric field and strong magnetic field, there exists a positive energy whereas for a low magnetic field, the energy of the hydrogen atom in quantum plasma becomes negative. From these figures, we observed that AB-flux field seems to be the key parameter. All these justify the superiority effect of AB flux over magnetic and electric fields on hydrogen atom in quantum plasmas. This can be understood further if we consider a strong electric field (i.e. $F=1.2$) and then calculate $\Delta E_{nm}=E_{nm}|_{B=5}-E_{nm}|_{B=0}$ for the three plots we discussed above. From (a), where $\xi=1$, $\Delta E_{nm}\approx38$. In (b) where $\xi=2$, $\Delta E_{nm}\approx160$. In subplot of (b), where $\xi=4$, $\Delta E_{nm}\approx2500$.

In Figure \ref{fig3}(c), we find that the external electric field will either has no effect on the energy of the hydrogen atom in quantum plasmas or will decrease the energy values under high intensity. It can be concluded from Figures \ref{fig3} and \ref{fig4}, that the confinement effect of AB-flux field on hydrogen atom dominates on external electric field and more dominance on magnetic field. Therefore,  AB-flux field can be regarded as a key control parameter for energy levels or localization of quantum state of hydrogen atom in quantum plasmas. In other word, to maintain a low-energy for hydrogen atom in quantum plasmas, a strong electric field and weak magnetic field are required where AB-flux field can serve as a regulator.

\section{Concluding remarks}
In this paper, we have studied the effects of electric field, AB-flux field and uniform magnetic field directed along $z$-axis  on hydrogen atom in quantum plasmas. The overall effects indicate that the system is strongly attractive while the localizations of quantum levels change and the eigenvalues decrease.  Also, as we have demonstrated, the combined effect of the fields is stronger than individual effects and consequently, there is a considerably shift in the bound state energy of the system. We found that to maintain a low-energy for hydrogen atom in quantum plasmas, a strong electric field and weak magnetic field are required where AB-flux field can be used as a regulator or a booster. The application of perturbation technique we utilized in this paper is not limited to plasma physics, it can also be applied in molecular physics.

Finally, we suggest a possible extension of the current work for inclusion of quantum effects with two and three- particle correlations. \cite{BJ21,BJ22}\\[1ex]

\begin{acknowledgments}
We thank the kind referees for the positive enlightening comments and suggestions, which have greatly helped us in making improvements to this paper. In addition, BJF acknowledges eJDS (ICTP) Prof. K. J. Oyewumi for his continous supports. This work is partially supported by 20150964-SIP-IPN, Mexico.
\end{acknowledgments}

\appendix

\section{Exact solution to a general potential form under the influence of AB-flux and external magnetic fields}
In this section, we show in detail, the derivation of equation (\ref{EE1}) for a general form of potential model $V(r)$, i.e. $\left(i\hbar\vec{\nabla}-{e}/{c}\vec{A}\right)^2\psi=2\mu(E_{nm}-V(r))\psi$. For convieniency, let us introduce $\mathcal{K}=-e/c$, so that equation (\ref{EE1}) becomes:
\begin{equation}
-\hbar^2{\nabla}^2\psi+i\hbar\mathcal{K}\vec{\nabla}\cdot(\vec{A}\psi)+i\hbar\mathcal{K}\vec{A}\cdot\vec{\nabla}\psi+\hbar\mathcal{K}^2\vec{A}\cdot\vec{A}\Psi=2\mu(E_{nm}-V(r))\psi.\tag{A1}
\label{A1}
\end{equation}
Using the property $\vec{\nabla}\cdot(\vec{A}\psi)=\vec{A}\cdot\vec{\nabla}\psi+\psi\vec{\nabla}\cdot\vec{A}$ and
\begin{equation}
 \vec{\nabla}\cdot\vec{A}=\left(\frac{\partial}{\partial r}\hat{r} +\frac{1}{r}\frac{\partial}{\partial\phi}\hat{\phi}+\frac{\partial}{\partial z}\hat{z}\right)\cdot\left(\frac{Br}{2}+\frac{\phi_{AB}}{2\pi r}\right)\hat{\phi}=\frac{1}{r}\frac{\partial}{\partial\phi}\vec{A}=0.\tag{A2}
\label{A2}
\end{equation}
Hence, equation (\ref{A1}) becomes
\begin{equation}
-\hbar^2{\nabla}^2\psi+2i\hbar\mathcal{T}\vec{A}\cdot\vec{\nabla}\psi+\hbar\mathcal{K}^2\vec{A}\cdot\vec{A}\Psi=2\mu(E_{nm}-V(r))\psi.\tag{A3}
\label{A3}
\end{equation}
Now, we obtain expression for $\nabla^2\psi$ and $\vec{A}\cdot\vec{\nabla}\psi$ as:
\begin{widetext}
\begin{align*}
&&\nabla^2\psi=\frac{1}{r}\frac{\partial}{\partial r}\left(r\frac{\partial\psi}{\partial r}\right)+\frac{1}{r^2}\frac{\partial^2\psi}{\partial\phi^2}+\frac{\partial^2\psi}{\partial z^2}={r^{-\frac{1}{2}}e^{im\phi}}\left(\frac{\mathcal{H}(r)}{4r^2}+\mathcal{H}''(r)-\frac{m^2}{r^2}\mathcal{H}(r)\right)\nonumber\\
&&\vec{A}\cdot\vec{\nabla}\psi=\left(\frac{Br}{2}+\frac{\phi_{AB}}{2\pi r}\right)\hat{\phi}\cdot\left(\frac{\partial\psi}{\partial r}\hat{r} +\frac{1}{r}\frac{\partial\psi}{\partial\phi}\hat{\phi}+\frac{\partial\psi}{\partial z}\hat{z}\right)=\frac{im}{r}\left(\frac{Br}{2}+\frac{\phi_{AB}}{2\pi r}\right)\tag{A4}
\label{A4}
\end{align*}
Substituting equation (\ref{A4}) into equation (\ref{A3}), we have
\begin{equation}
-\hbar^2\mathcal{H}''(r)-\frac{\hbar^2\mathcal{H}(r)}{4r^2}+\frac{m^2\mathcal{H}(r)\hbar^2}{r^2}-\frac{2\hbar m\mathcal{K}}{r}\left(\frac{Br}{2}+\frac{\phi_{AB}}{2\pi r}\right)\mathcal{H}(r)+\mathcal{K}^2\vec{A}\cdot\vec{A}\mathcal{H}(r)=2\mu(E_{nm}-V(r))\mathcal{H}(r).\tag{A5}
\end{equation}
Using $e=c=1$, then $\mathcal{K}=-1$, so  we obtain a more explicit expression as:
\begin{equation}
\mathcal{H}''(r)+\frac{2\mu}{\hbar^2}\left[E-\left(V(r)-Fr+\frac{\hbar^2}{2\mu}\left(\frac{(m+\xi)^2-\frac{1}{4}}{r^2}\right)+\frac{\omega_c\hbar}{2}(m+\xi)+\left(\frac{\mu\omega_c^2}{8}\right)r^2\right)\right]\mathcal{H}(r)=0, \tag{A6}
\end{equation}
where we have introduced $\xi=\phi_{AB}/\phi_0$ with $\phi_0=hc/e$ and electric field $F$.
\end{widetext}


\end{document}